\documentclass{emulateapj}
\usepackage{apjfonts}
\usepackage{amsmath}

\newcommand{\myemail}{kmurase@yukawa.kyoto-u.ac.jp}

\def\mr#1{\mathrm{#1}}

\newcounter{ichi}
\newcounter{ni}
\newcounter{san}
\slugcomment{Submitted}

\shorttitle{Probing Intergalactic Magnetic Fields in the GLAST Era}
\shortauthors{Murase et al.}

\begin{document}


\title{Probing Intergalactic Magnetic Fields in the GLAST Era\\
through Pair Echo Emission from TeV Blazars}


\author{
Kohta Murase\altaffilmark{1}, 
Keitaro Takahashi\altaffilmark{1},
Susumu Inoue\altaffilmark{2}, 
Kiyomoto Ichiki\altaffilmark{3},
and Shigehiro Nagataki\altaffilmark{1}
}


\altaffiltext{1}{YITP, Kyoto University, Kyoto, Oiwake-cho,
Kitashirakawa, Sakyo-ku, Kyoto 606-8502, Japan; \myemail}
\altaffiltext{2}{Department of Physics, Kyoto University, Kyoto, Oiwake-cho,
Kitashirakawa, Sakyo-ku, Kyoto 606-8502, Japan}
\altaffiltext{3}{Research Center for the Early Universe, University of
Tokyo, 7-3-1 Hongo, Bunkyo-ku, Tokyo 113-0033, Japan}


\begin{abstract}
More than a dozen blazars are known to be emitters of multi-TeV gamma rays,
often with strong and rapid flaring activity.
By interacting with photons of the cosmic microwave and infrared backgrounds,
these gamma rays inevitably produce electron-positron pairs,
which in turn radiate secondary inverse Compton gamma rays in the GeV-TeV range
with a characteristic time delay that depends on the properties of  
the intergalactic magnetic field (IGMF).
For sufficiently weak IGMF, such ``pair echo'' emission may be detectable
by the Gamma-ray Large Area Space Telescope (GLAST),
providing valuable information on the IGMF.
We perform detailed calculations of the time-dependent spectra of pair echos
from flaring TeV blazars such as Mrk 501 and PKS 2155-304,
taking proper account of the echo geometry and other crucial effects.
In some cases, the presence of a weak but non-zero IGMF may enhance the detectability of echos.
We discuss the quantitative constraints that can be imposed on the IGMF
from GLAST observations, including the case of non-detections.    
\end{abstract}





\section{Introduction}
Magnetic fields are observed to be common in the structured regions of the universe,
such as galaxies and clusters of galaxies.
To interpret their microgauss field strengths,
the pre-collapse seed fields in the intergalactic medium
required by dynamo theories may be of order ${10}^{-20}$ G \cite{Kul1}.
Various theoretical possibilities have been suggested
for the origin of such large-scale, intergalactic magnetic fields (IGMFs).
For example, primordial magnetic fields of $\sim {10}^{-20}$ G
may be generated during the cosmological QCD phase transition \cite{Sig1}.
They can also be produced during the inflationary epoch
if conformal invariance of electromagnetic interactions is broken \cite{Tur1}.
Density fluctuations before cosmic recombination 
inevitably give rise to weak IGMFs \cite{Tak1,Ich1}. 
The Biermann battery mechanism \cite{Gne1}
or radiation drag effects \cite{Lan1} at cosmic reionization fronts 
can induce fields of ${10}^{-20} - {10}^{-16}$ G.
At low redshift, such weak IGMFs are expected to survive
in intergalactic void regions, as they can remain uncontaminated
by astrophysical sources such as galactic winds or quasar outflows \cite{Fur1}.
Measurements of IGMFs would be crucial for understanding the origin of galactic magnetic fields.

One of the best-known tools to probe magnetic fields is Faraday rotation measurements,
from which an upper limit of $\sim {10}^{-9}$ G has been derived
for the IGMF assuming correlation lengths $\sim 1$ Mpc \cite{Kro1}. 
A different approach suitable for probing very weak IGMFs is to utilize ``pair echo'' emission
from transient very-high-energy (VHE) gamma-ray emitters such
as blazars and gamma-ray bursts (GRBs) \cite{Pla1}.
Multi-TeV photons from distant sources are attenuated
by $\gamma \gamma$ pair-production interactions with the cosmic microwave background 
(CMB) and cosmic infrared background (CIB). The electron-positron pairs
then up-scatter CMB and CIB photons by the inverse-Compton (IC) 
process to produce secondary gamma rays, whose flux depends strongly 
on the IGMF. If it is stronger than $\sim {10}^{-12}$ G,
the formation of a very extended and nearly isotropic pair halo is unavoidable \cite{Aha1}.
On the other hand, if the IGMF is sufficiently weak, most of the secondary gamma rays
will come from the direction of the source with some temporal and
spatial spreading as pair echos, containing valuable information on the IGMF. 
Blazars are promising sources for this purpose,
since they are observed to exhibit strong flares at multi-TeV energies
with flux variations by a factor of $\gtrsim 10$ over time scales of $\lesssim 1$ hr to months. 
Pair echo emission is typically expected at GeV energies,
appropriate for the recently launched Gamma-ray Large Area Space Telescope (GLAST) satellite.

Here we reconsider pair echo emission from blazars by 
exploiting the formulation recently developed by Ichiki et al. (2008; see also Takahashi et al. 2008),
which allows more satisfactory calculations of the time-dependent echo
spectra compared to previous works \cite{Dai1,Fan1}.   
Since the pair echo emission can persist after the primary flare decays, it may 
be observable unless it is hidden by some quiescent emission. Furthermore,
constraints on the IGMF are possible even in the case of GLAST non-detections.
We demonstrate this using as examples
the past flares from Mrk 501 in 2005 and from PKS 2155-304 in 2006.
Further details of our methods and results will be described in a subsequent paper
(Takahashi et al, in preparation).


\section{\label{sec:1}Emission Properties}
So far, 23 AGNs have been detected at energies $\geq 0.1$ TeV
\footnote{http://www.mppmu.mpg.de/~rwagner/sources/} (Wagner 2008). 
Most of them belong to the high-frequency-peaked BL
Lac subclass of blazars, characterized by spectral energy
distributions with two maxima occurring in the X-ray and TeV gamma ray bands.
The standard blazar model comprises supermassive black holes
ejecting relativistic jets close to the line of sight.
Owing to relativistic beaming,
blazars exhibit rapid flux variations on timescales down to a few minutes,
with strong TeV flares being the most extreme events.

The $\gamma \gamma$ optical depth of the CIB depends on 
gamma-ray energy, source redshift and CIB intensity.
For the CIB, we adopt here the ``best-fit'' and ``low-IR'' models
of Kneiske et al. (2002, 2004) \citep[see also][]{Pri1,Ste3}.
Note that 
recent observations of TeV blazars may point to a CIB resembling the low-IR model,
close to the lower limits from galaxy count data (Aharonian
et al. 2006; Albert et al. 2008; see however, Stecker et al. 2007).

\subsection{\label{subsec:a}Pair Echo Emission}
The total fluence of pair echo emission is determined by
the $\gamma \gamma$ optical depth of the CIB and does not depend on the IGMF.
Primary photons with energy $E_{\gamma}$ 
are converted to electron-positron pairs with Lorentz factor 
$\gamma_e \approx {10}^{6} (E_{\gamma}/1 \, {\rm{TeV}}) (1+z)$ 
in the local cosmological rest frame,
which then up-scatter CMB and CIB photons.
CMB photons are boosted to energies
$\sim 2.82 k_B T_{\rm{CMB}}^{\prime} \gamma_e^2 /(1+z)
\approx 0.63 {(E_{\gamma}/1 \, \rm{TeV})}^{2} {(1+z)}^{2}$ GeV,
where $T_{\rm{CMB}}^{\prime} \simeq 2.73 (1+z)$ K is the local CMB temperature.
To evaluate the pair echo flux,
we must consider various time scales involved in the process,
such as the flare duration, angular spreading time,
and the delay time due to magnetic deflections \cite{Dai1,Fan1,KM4}. 
These can be estimated as follows. 

The angular spreading time is
${\Delta t}_{\mr{ang}} \approx (1+z) (\lambda _{\rm IC}^{\prime} + \lambda
_{\gamma \gamma}^{\prime})/2{{\gamma}_{e}}^{2}c$,
where $\lambda _{\gamma \gamma}^{\prime} \approx  
{(0.26 \sigma_T n'_{\rm{CIB}})}^{-1} \approx 20 \, {\rm Mpc} {(n'_{\rm{CIB}}/ 0.1 \, {\rm{cm}^{-3}})}^{-1}$
is the local $\gamma\gamma$ mean free path
in terms of the local CIB photon density $n'_{\rm{CIB}}$,
and $\lambda _{\rm IC}^{\prime}=3 m_e c^2/(4 \sigma_T U'_{\rm{CMB}} \gamma_e)
\approx 690 \, {\rm kpc} {(\gamma_e/{10}^{6})}^{-1} {(1+z)}^{-4}$
is the local IC cooling length
in term of the local CMB energy density $U'_{\rm{CMB}}$.
At the energies of our interest, $\lambda' _{\gamma \gamma} \gg \lambda' _{\rm IC}$ so that
${\Delta t}_{\mr{ang}} \approx (1+z) 
\lambda _{\gamma \gamma}^{\prime}/2{{\gamma}_{e}}^{2}c
\approx 960 \, {\rm{s}} {(\gamma_e/{10}^{6})}^{-2}
{(n'_{\rm{CIB}}/ 0.1 \, {\rm{cm}^{-3}})}^{-1} (1+z)$.
For sufficiently small deflections in weak IGMFs
with present-day amplitude $B_{\rm IG}=B_{\rm IG}^{\prime} {(1+z)}^{-2}$ and 
coherence length $\lambda_{\rm{coh}}=\lambda_{\rm{coh}}^{\prime} (1+z)$,
the magnetic deflection angle is
$\theta_B = {\rm{max}}[\lambda_{\rm{IC}}^{\prime}/r_L,
({\lambda_{\rm{IC}}^{\prime} 
{\lambda}_{\rm{coh}}^{\prime})}^{1/2}/r_L]$,
where  $r_{\mr{L}}=\gamma_e m_e c^2/e B_{\rm IG}^{\prime}$ is the
Larmor radius of the electrons or positrons.
The delay time due to magnetic deflections is
${\Delta t}_{B} \approx (1+z) (\lambda _{\rm IC}^{\prime}+ 
\lambda _{\gamma \gamma}^{\prime}) ({\theta}_{B}^{2}/2c)$.
Note that prior to Ichiki et al. (2008), the $\lambda'_{\gamma \gamma}$ 
term here had been neglected and ${\Delta t}_{B}$ 
underestimated by up to 2-3 orders of magnitude.
For coherent magnetic fields with
$\lambda'_{\rm{coh}} \gtrsim \lambda'_{\rm{IC}}$,
we have ${\Delta t}_{B} \approx {\rm{max}}
[6.1 \times {10}^{3} \, {\rm{s}} {(\gamma_e/{10}^{6})}^{-5}
{(B_{\rm{IG}}/ {10}^{-20} \, \rm{G})}^{2} {(1+z)}^{-7}, 
1.6 \times {10}^{5} \, {\rm{s}} {(\gamma_e/{10}^{6})}^{-4} 
{(n_{\rm{CIB}}/ 0.1 \, {\rm{cm}^{-3}})}^{-1} 
{(B_{\rm{IG}}/ {10}^{-20} \, \rm{G})}^{2} {(1+z)}^{-3}]$.
Implicit in the above discussion is that both $1/\gamma_e$ and 
$\theta_B$ do not exceed $\theta_j$, the opening angle of the AGN jet;
otherwise a significant fraction of photons or pairs will be
deflected out of the line of sight and the echo greatly diminished.

Together with the flare duration $T$, the pair echo delay time can be estimated by
$\Delta t= {\rm{max}}[{\Delta t}_{\mr{ang}}, {\Delta t}_{B}, T]$.
If ${\Delta t}_B$ dominates, pair echos can serve as effective probes of weak IGMFs.
For flaring blazars with $T \sim$ day, ${\Delta t}_{\mr{ang}}$ is irrelevant,
and ${\Delta t}_B \gtrsim T$ at echo emission energies
$\lesssim 0.86 \, {\rm{GeV}} (B_{\rm{IG}}/{10}^{-20} \, {\rm{G}}) 
{(T/1 \, {\rm{day}})}^{-1/2} {(n_{\rm CIB}/0.1 \, \rm{cm}^{-3})}^{-1/2} 
{(1+z)}^{-3/2}$ for the case of coherent IGMFs.
Such estimates were the basis of previous evaluations of the pair echo flux \cite{Dai1,Fan1},
but explicit descriptions of the time-dependent spectra were not possible without some ad hoc modifications \cite{KM4}.
In constrast, the formulation of Ichiki et al. (2008)
enables us to calculate the time-dependent spectra in a more satisfactory manner, particularly at late times,
accounting properly for the geometry of the pair echo process. 

\subsection{\label{subsec:b}Primary Emission Spectrum}
First we consider the archetypal flaring blazar Mrk 501.
Strong flares at energies up to 20 TeV were observed in 1997 by HEGRA
(e.g., Katarzy\'nski et al. 2002).
Similar strong flares were recently observed by MAGIC from May through July 2005 \cite{Alb1},
during which the flux varied by an order of magnitude, and intranight variability was observed
with flux-doubling times down to 2 minutes on the nights of June 30 and July 9.
We focus on the strong flare of June 30.

The second example is PKS 2155-304, the brightest VHE blazar
in the southern hemisphere \cite{Aha3,Aha4}.
During July 2006, the average VHE flux observed by HESS was
more than 10 times its quiescent value in 2003 \cite{Aha6}.
In particular, an extremely strong flare was observed on July 28, 
$\sim 50$ times brighter than the quiescent level,
which we use as a template for calculating the pair echo emission. 
Only small spectral differences were found between the flaring and quiescent states,
as opposed to other blazars that often reveal large spectral changes at different flux levels \cite{Aha6}.

From a theoretical viewpoint,
the intrinsic maximum energy $E_{\gamma}^{\rm{max}}$
should reflect either the maximum energy of accelerated electrons or protons,
or a cutoff due to internal $\gamma \gamma$ absorption.
Although the true value of $E_{\gamma}^{\rm{max}}$ is not yet known,
the spectrum of Mrk 501 was observed to extend at least to $\sim 20$ TeV,
so here we take $E_{\gamma}^{\rm{max}}=20$ TeV as a reasonable assumption.
Concerning the emission mechanism,
both leptonic and hadronic models have been proposed
(see, e.g. Sikora \& Madejski 2001; B\"ottcher 2006 for reviews).
The synchrotron self Compton (SSC) model is one of the
most popular leptonic models (e.g., Maraschi et al. 1992; Inoue \& Takahara 1996).
Another frequently discussed leptonic model is the external Compton model,
where electrons up-scatter external photons originating outside the jet.
In BL Lac objects, the tightly correlated variability in the X-ray and TeV bands
(e.g., Katarzy\'nski et al. 2005) and the lack of strong emission lines
indicate a minor role for external photons and favor the SSC model.
In hadronic models involving protons accelerated to ultra-high energies,
the high-energy spectra are attributed to synchrotron radiation from either
the protons themselves (Aharonian 2000),
or secondary electron-positron pairs generated in photohadronic interactions (Mannheim 1993).
The hadronic models are challenged by the observed X-ray -TeV
correlation and rapid gamma-ray variability,
but they have not been entirely ruled out.

\section{\label{sec:2}Results}
\begin{figure}[tb]
\includegraphics[width=\linewidth]{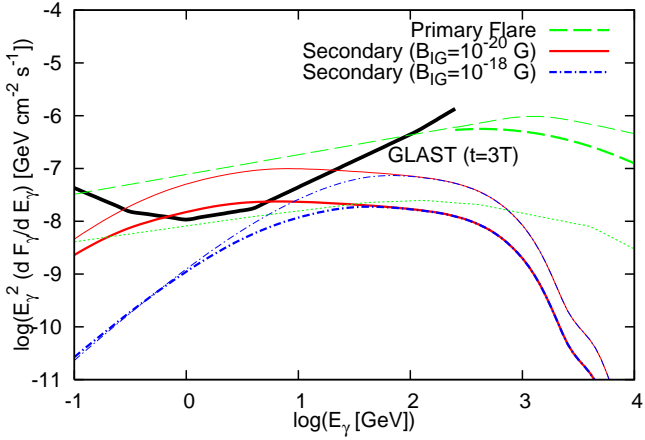}
\caption{\label{f1} 
Pair echo spectra for the 2005 flare of Mrk 501,
plotted at $t=T=0.5$ day (thin) and $t=3 T=1.5$ day (thick),
for the cases of $B_{\rm IG}=10^{-20}$ G (solid) and $10^{-18}$ G (dot-dashed)
with $\lambda_{\rm{coh}}=0.1$ kpc.
Also shown are the observed primary spectrum (thick dashed)
and intrinsic primary spectrum for the low-IR CIB model (thin dashed) at $t=0$,
described by linear extrapolation at $\lesssim 200$ GeV.
The quiescent emission is represented by an SSC model (dotted).
The GLAST sensitivity for integration time $t=3 T=1.5$ day is overlayed.}
\end{figure}
\begin{figure}[tb]
\includegraphics[width=\linewidth]{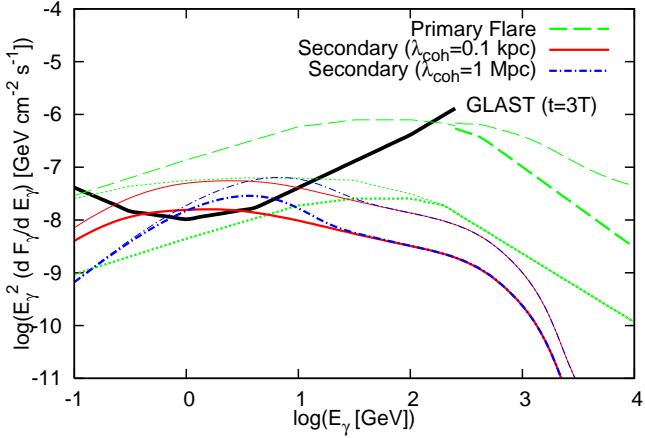}
\caption{\label{f2}
Pair echo spectra for the 2006 flare of PKS 2155-304,
plotted at $t=T=0.5$ day (thin) and $t=3 T=1.5$ day (thick),
for the cases of $\lambda_{\rm{coh}}=0.1$ kpc (solid) and $\lambda_{\rm{coh}}=1$ Mpc (dot-dashed),
with $B_{\rm IG}=10^{-20}$ G.
Also shown are the observed primary spectrum (thick dashed)
and the intrinsic primary spectrum for the low-IR CIB model (thin dashed) at $t=0$,
described by a hadronic model at $\lesssim 200$ GeV.
The quiescent emission is represented by an SSC (thin dotted) or a hadronic (thick dotted) model.
The GLAST sensitivity for integration time $t=3 T=1.5$ day is overlayed.}
\end{figure}
\begin{figure}[tb]
\includegraphics[width=\linewidth]{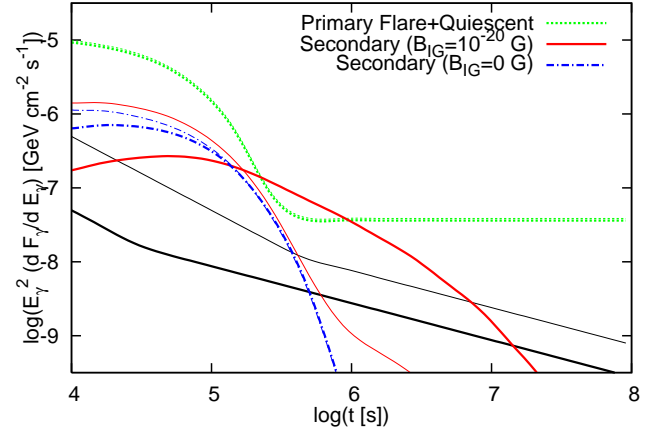}
\caption{\label{f3} 
Primary and pair echo light curves for a hypothetical strong flare of PKS 2155-304
compared with the GLAST sensitivity at 1 GeV (thick) and 10 GeV (thin),
for the case of $B_{\rm IG}=10^{-20}$ G (solid) and $B_{\rm IG}=0$ G (dot-dashed)
with $\lambda_{\rm{coh}}=1$ Mpc. The best fit CIB model is employed.
Also shown is the total primary light curve
for both flare and quiescent emission assuming an SSC model.}
\end{figure}
\begin{table}[t]
\begin{center}
\begin{tabular}{|c|c|c|c|c|}
\hline Flare Source & CIB Model & Duration $T^{\prime}$&
Expected Lower Limit $B_{\rm{IG}} {\lambda}_{\rm{coh}}^{1/2}$\\
\hline
\hline Mrk 501 & low-IR & 0.5 day & ${10}^{-20.5}$ $\rm{G} \, 
\rm{Mpc}^{1/2}$\\
\hline Mrk 501 & best-fit & 0.5 day & ${10}^{-19}$ $\rm{G} \, \rm{Mpc}^{1/2}$\\
\hline PKS 2155-304 & low-IR & 0.5 day & ${10}^{-21.5}$ $\rm{G}
\, \rm{Mpc}^{1/2}$\\
\hline PKS 2155-304 & best-fit & 0.5 day & ${10}^{-20.5}$ $\rm{G} \, 
\rm{Mpc}^{1/2}$\\

\hline
\end{tabular}
\caption{Lower bounds on the IGMF in the case of GLAST non-detections of pair echos
\label{LP}}
\end{center}
\end{table}


Fig. 1 shows the time-dependent pair echo spectra for the 2005 flare of Mrk 501
at different times $t$ after the onset of the flare,
whose flux is assumed to decay as $\exp(-t/T)$ on a timescale $T'=T/(1+z)=0.5$ day.
We take $\lambda_{\rm{coh}}=0.1$ kpc and different values of $B_{\rm IG}$.
The sub-TeV primary spectrum of Mrk 501 is relatively hard
with photon index  $\alpha \lesssim 0.4$ \cite{Aha2,Alb1}.
Extrapolating this to GeV, the pair echo should be visible relative to the primary flare.
It should not be masked by the quiescent GeV emission either,
if the latter is estimated by a one-zone SSC model 
consistent with the ``low flux'' TeV data of 2005 as in Fig. 21 of Albert et al. (2007).
Compared with the GLAST sensitivity \footnote{http://www-glast.stanford.edu},
the detection prospects for pair echos seem reasonable
as long as the IGMF is sufficiently weak (see also Dai et al. 2002).
Detailed observations of its spectra and light curve
will provide valuable information on the IGMF (Takahashi et al., in preparation).

Note, however, that Mrk 501 has been previously detected at GeV by EGRET
during a multiwavelength campaign in March 1996 \cite{Kat1}.
The TeV spectrum was much harder than in 1997, and the entire GeV-TeV spectrum
was incompatible with the simplest, one-zone SSC model.
If this GeV emission corresponds to a persistent, quiescent component
of Mrk 501, it may obscure the pair echo emission.
This question should be resolved soon by GLAST.

Results for the 2006 flare of PKS 2155-304 are shown in Fig. 2,
with assumptions for the flare similar to Mrk 501.
Here we fix $B_{\rm{IG}}={10}^{-20}$ G and show the dependence on $\lambda_{\rm{coh}}$.
The case of $\lambda_{\rm{coh}}=0.1$ kpc corresponds to tangled IGMFs
($\lambda_{\rm{coh}} \lesssim \lambda_{\rm{IC}}$),
while $\lambda_{\rm{coh}}=1$ Mpc implies coherent IGMFs
($\lambda_{\rm{coh}} \gtrsim \lambda_{\rm{IC}}$).
The quiescent GeV emission is estimated with both SSC and hadronic models
for the average TeV flux in 2003 as in Fig. 10 of Aharonian et al. (2005b). 
The flare spectrum of PKS 2155-304 above TeV is much softer than in Mrk 501 \cite{Aha4,Aha6}.
This may also be true below TeV if the SSC model is adopted \cite{Aha5},
in which case the pair echo may be hidden beneath the primary and quiescent emission.

Generally speaking, stronger IGMFs dilute the echo emission
and make its observation more difficult.
However, in some cases, the presence of the IGMF can enhance
the detectability of the echo emission, at least in principle.
As an illustrative example, we show the results for a hypothetical TeV flare
of PKS 2155-304 that is a further 10 times stronger than in Fig. 2,
using the best fit CIB model.
Here the pair echo emission below $\sim$ GeV 
outlasts the primary flare by virtue of the IGMF,
enabling the echo to emerge after the flare subsides.

Next we discuss the lower limits that can be imposed on the IGMF
even when GLAST does not detect pair echos from TeV blazars.
Such limits would be valid if the primary flare and quiescent emission 
at GeV energies are low enough so that the case of $B_{\rm{IG}}=0$
implies an excess of the echo flux $dF_{\gamma,\rm{sec}}/d E_{\gamma}$
over the primary flux $dF_{\gamma,\rm{pri}}/d E_{\gamma}$.
The non-detection of the echo emission can then be attributed to the effects of a finite IGMF,
expressed as
$(dF_{\gamma,\rm{sec}}/d E_{\gamma}) < {\rm{max}}
[(d F_{\gamma,\rm{pri}}/ d E_{\gamma}), 
(d F_{\gamma,\rm{lim}}/d E_{\gamma})]$, where 
$dF_{\gamma,\rm{lim}}/d E_{\gamma}$ is the GLAST sensitivity. 
Summarized in Table 1 are the constraints thus derived for the IGMF,
which depend on the CIB as well as on assumptions for the primary emission.
For PKS 2155-304, the limits are given only for the hadronic model,
since the $B_{\rm{IG}}=0$ echo flux is not expected
to exceed the primary flux for the SSC model.
More conservative but less model-dependent constraints may be deduced
for blazars such as Mrk 501 with hard primary TeV spectra,
where the primary GeV emission is expected to be less obstructive (Fig. 1).
Note that these limits are for tangled IGMF that lead to
lower echo fluxes at $\lesssim 10$ GeV and hence more conservative limits
compared to coherent fields.
The same is true for low-IR models compared to best-fit models for the CIB.
A more detailed account of the primary light curve should
allow more realistic constraints.

\section{\label{sec:3}Summary and Discussion}
We have evaluated the time-dependent spectra of secondary pair echo emission from TeV blazars
and discussed the information that can be derived for the IGMF,
applying a recently developed formalism of pair echos that properly describes their time evolution.
The observational prospects are quite interesting for the recently launched GLAST mission,
and successful detections would open a new window on studies of cosmic magnetic fields.
Even in the case of non-detections,
lower limits on the IGMF of $B_{\rm{IG}} {\lambda}_{\rm{coh}}^{1/2}  
\gtrsim ({10}^{-19} - {10}^{-21}) \, \rm{G} \, \rm{Mpc}^{1/2}$
may be obtained, making use of suitably strong TeV flares with hard spectra.
The existence of a weak but non-zero IGMF may also sometimes enhance
rather than diminish the detectability of echo emission at late times.

Further detailed calculations utilizing Monte Carlo methods
may be desirable for more robust predictions.
For example, the effect of cooling of pairs during propagation in the IGMF
can be moderately important, leading to fluxes smaller by a factor of several
compared to the results given here \cite{KM4}.
Also of concern are uncertainties in the CIB models,
which can affect not only the pair echo fluence
but also the timescales for angular spreading and magnetic deflection delay at all redshifts.
We must also beware of uncertainties in the intrinsic primary spectra
including the value of $E_{\gamma}^{\rm{max}}$.


In addition to pair echos from flares, the quiescent emission of TeV blazars
may also contain useful information on the IGMF.
For example, depending on the CIB,
the spectra corrected for $\gamma \gamma$ absorption
for some blazars including Mrk 501
point to a sharp pile-up at high energies,
contradicting the expectations from conventional emission models.
However,
as long as $B_{\rm{IG}} \lesssim {10}^{-18}$ G
and the primary spectra has photon index $\alpha \sim 2$ and $E_{\gamma}^{\rm{max}} \gtrsim 100$ TeV,
an intergalactic cascade component may contribute to the quiescent TeV emission
and compensate the effects of CIB absorption \cite{Aha2}.
The quiescent flux at $\lesssim$ GeV could also be affected by such cascades \cite{Dai1}. 
These issues will be elaborated on in a future publication (Takahashi et al., in preparation).



\acknowledgments
K.M., K.T. and K.I. are supported by JSPS fellowships.
S.I. is supported by Grants-in-Aid for Scientific Research
of the Ministry of Education, Culture, Science, Sports and Technology of Japan (MEXT),
Nos. 19047004 and 19540283.
S.N. is supported likewise by Nos. 19104006, 19740139, 19047004.
Support also comes from the Grant-in-Aid for the Global COE Program 
"The Next Generation of Physics, Spun from Universality and Emergence" from MEXT.

\clearpage





\end{document}